\def\o{\over}
\def\Ar{\rightarrow}
\def\bar{\overline}

\def\a{\alpha}

\def\n{\nu}
\def\m{\mu}

\def\e{\epsilon}

\def\th{\theta}

\def\bar{\overline}
\def\l{\lambda}

\def\eV{{\rm eV}}

\documentstyle [12pt,epsf]{article}
\begin{document}
\baselineskip=24pt
\setcounter{page}{1}
\thispagestyle{empty}
\topskip 2.5  cm
\topskip 0.5  cm
\begin{flushright}
\begin{tabular}{c c}
& {\normalsize hep-ph/9807283v2}\\
& EHU-07 Revised, September  1998
\end{tabular}
\end{flushright}
\vspace{1 cm}
\centerline{\LARGE\bf Vacuum Neutrino Oscillations of Solar}
 \centerline{\LARGE\bf  Neutrinos and Lepton Mass Matrices}

\vskip 1.5 cm
\centerline{{\large \bf Morimitsu TANIMOTO}
  \footnote{E-mail address: tanimoto@edserv.ed.ehime-u.ac.jp}
   }
\vskip 0.8 cm
 \centerline{ \it{Science Education Laboratory, Ehime University, 
 790-8577 Matsuyama, JAPAN}}
\vskip 3 cm
\centerline{\bf ABSTRACT}\par
\vskip 0.5 cm
   We consider the case that the solar neutrino deficit is due to the vacuum oscillation.
   The lepton mass matrices with nearly bi-maximal mixings are needed 
   in order to explain both solar and  atmospheric neutrino deficit.
     A texture with the symmetry of flavour democracy or $S_3$
	  has been investigated by taking account of the symmetry breaking terms of 
	  the charged lepton mass matrix.
	It is found that predicted mixings can be considerably changed from
	 the neutrino mixings $\sin^2 2\th_\odot\simeq  1$ and
	  $\sin^2 2\th_{\rm atm}\simeq 8/9$ at the symmetric limit.
	  The correlation between $|U_{e3}|$ and $|U_{e1}U_{e2}^*|$ is also presented.
  The test of the model is discussed by focusing on the three flavor analyses
   in the solar neutrinos, atmospheric neutrinos and long baseline experiments.
  
\vskip 0.5 cm
 \newpage
%%%%%%%%%%%%%%%%%%%%%%%%%%%%%%%%%%%%%%%%%%%%%%%%%%%%%%%%%%%%%%%%%%%%%%%%%%%%%%%
%%%%%%%%%%%%%%%%%%%%%%%%%%%%%%%%%%%%%%%%%%%%%%%%%%%%%%%%%%%%%%%%%%%%%%%%%%%%%%%
\topskip 0.  cm

 Neutrino flavor oscillations provide 
  information of the fundamental  property of neutrinos  such as  masses, 
flavor mixings.  
  In these years, there is growing  experimental evidences of  neutrino
oscillations.
  The exciting one is the atmospheric neutrino deficit \cite{Atm1}$\sim$\cite{Atm3}
    as well as the solar neutrino deficit \cite{solar}.
	   Super-Kamiokande \cite{SKam}  presented the near-maximal
     neutrino flavor oscillation in  atmospheric neutrinos.
  Recent Super-Kamiokande data also suggest that vacuum oscillation regions are favored
   in the analysis of day-night spectra and energy shape of the solar neutrino 
   \cite{SKamsolar}$\sim$\cite{Goldhaber}. 
    Therefore, three flavor analyses are very important for the vacuum oscillation of 
 solar neutrinos \cite{Smir} \cite{Osland}.
	Results of those analyses can give constraints on the structure of the lepton mass matrices.
  Therefore, it is urgent to make clear predictions quantitatively in  
   the lepton mass matrix models  which give  near-maximal solar and
    atmospheric vacuum oscillations \cite{Barger}$\sim$\cite{Nomura}.

	There is a typical texture of the lepton mass matrix with
	 nearly bi-maximal mixing of three neutrinos, which
	 is derived from the symmetry of the lepton flavor democracy
	  by Fritzsch and Xing \cite{FX}, or  from the $S_{3L}\times S_{3R}$ symmetry 
	  of the left-handed Majorana 
	neutrino mass matrix  given by  Fukugita, Tanimoto and Yanagida \cite{FTY}. 
	  These models  can give identical predictions for the neutrino mixings
	  $\sin^2 2\th_\odot\simeq 1$ and $\sin^2 2\th_{\rm atm}\simeq 8/9$
	  although those depend on  the symmetry breaking patterns.
	  However, these predictions are considerably  changed if the symmetry breaking terms
	   are correctly taking into account.
   In this paper, we investigate the quantitative predictions on the neutrino
	  flavor mixings including symmetry breaking terms,  which become
	  significant for long baseline(LBL) neutrino oscillation experiments.
	  In particular,  LBL experiments of $\n_\m\Ar \n_e$ 
	  can test the prediction of $U_{e3}$  in the  model.
  The first LBL  reactor experiment CHOOZ has already
 reported a  bound  of the neutrino oscillation \cite{CHOOZ},
    which gives a strong constraint of the flavor mixing pattern.
 The LBL accelerator experiment K2K \cite{K2K} is planned to begin taking data
 in the next year,  whereas the MINOS \cite{MINOS} and ICARUS \cite{ICARUS}
  experiments will start in the first year of the next century.
  Thus,  the lepton mass matrix model will be tested  in the near future.
  
%%%%%%%%%%%%%%%%%%%%%%%%%%%%%%%%%%%%%%%%%
%%%%%%%%%%%%%%%%%%%%%%%%%%%%%%%%%%%%%%%%%
 Our conservative approach is to assume that oscillations need only account for 
 the solar and atmospheric neutrino data.
 Since the result of LSND \cite{LSND} awaits confirmation by KARMEN experiment\cite{KARMEN},
  we do not take into consideration the LSND data in this paper.
  Recent results of atmospheric neutrinos
	  at Super-Kamiokande \cite{SKam}
  suggest $\nu_\m \Ar \nu_\tau$ oscillation with the near-maximal mixing.
 Since the CHOOZ result \cite{CHOOZ}
 excludes the large neutrino oscillation of $\nu_\mu \Ar \nu_e$ as far as
 $\Delta m^2 \geq 9\times 10^{-4} \eV^2$, 
 the large mixing between  $\nu_\m$ and $\nu_\tau$  is a reasonable interpretation
  for the atmospheric $\nu_\mu$ deficit.
  Our starting point as to neutrino mixings is
   the near-maximal $\nu_\m \Ar \nu_\tau$ oscillation with 
 \begin{equation}
  \Delta m^2_{\rm atm}=  10^{-3}\sim 10^{-2} \eV^2 \ , \qquad\quad
 \sin^2 2\th_{\rm atm} \geq 0.8 \ ,
  \label{atm}
 \end{equation}
\noindent which
  constrain   neutrino oscillations in  LBL experiments.
On the other hand, recent solar neutrino data of Super-Kamiokande \cite{SKamsolar} 
favors vacuum long-wavelength oscillations with $\Delta m_\odot^2 \simeq 10^{-10}\eV^2$ 
 and the  near-maximal mixing in the analysis of day-night spectra and energy shape.

%%%%%%%%%%%%%%%%%%%%%%%%%%%%%%%%%%%%%%%%%%%%%%
%%%%%%%%%%%%%%%%%%%%%%%%%%%%%%%%%%%%%%%%%%%%%%
%%%%%%%%%%     Model     %%%%%%%%%%%%%%%%%%%%%
%%%%%%%%%%%%%%%%%%%%%%%%%%%%%%%%%%%%%%%%%%%%%%

The texture of the lepton mass matrices with nearly bi-maximal mixing of three neutrinos
was presented based on the democratic mass matrix as follows \cite{FX}\cite{FTY}:
the charged lepton mass matrix is 
%%%%%%%%%%%%%%%%%%%%%%%%%%%%%%%%%%%%%%%%%%
%%%%%%%%%%%% Charged Leptons %%%%%%%%%%%%%%%%%%%%
%%%%%%%%%%%%%%%%%%%%%%%%%%%%%%%%%%%%%%%%%%
\begin{equation}
M_\ell= {c_\ell \over 3}
             \left (\matrix{1 & 1 & 1 \cr
			                1 & 1 & 1 \cr
                            1 & 1 & 1 \cr  } \right)
+\left( \matrix{\delta_\ell & 0 & 0 \cr
                         0 & \rho_\ell & 0 \cr
                         0 & 0 & \e_\ell \cr} \right) \ ,
\label{MC}
\end{equation}
\noindent where the second matrix is the symmetry breaking terms,
 which were given for quark mass matrices by Koide \cite{Demo}, and  $\delta_\ell$,
$\rho_\ell$ and $\e_\ell$ are complex parameters in general.
By neglecting $CP$ violating phases and using $c_\ell\gg \e_\ell \gg \rho_\ell, \delta_\ell$,
this matrix is diagonalized approximately  as
\begin{equation}
V_\ell^\dagger M_\ell V_\ell = {\rm diag}(m_{\ell 1}, m_{\ell 2}, m_{\ell 3}) \ , 
\end{equation}
where
\begin{eqnarray}
m_{\ell 1}&=&(\delta_\ell+\rho_\ell+\e_\ell)/3-\xi_\ell/6 \ ,\nonumber\\
m_{\ell 2}&=&(\delta_\ell+\rho_\ell+\e_\ell)/3+\xi_\ell/6 \ , \nonumber \\
m_{\ell 3}&=& c_\ell+(\delta_\ell+\rho_\ell+\e_\ell)/3 \ ,
\label{mass}
\end{eqnarray}
with
\begin{equation}
\xi_\ell=[(2\e_\ell-\rho_\ell-\delta_\ell)^2+3(\rho_\ell-\e_\ell)^2 ]^{1/2} \ . 
\end{equation}												
\noindent
The unitary matrix $V_\ell$ is given as $V_\ell=F L$, where
\begin{eqnarray}
F=&& \left( \matrix{1/\sqrt 2 & 1/\sqrt 6 & 1/\sqrt 3 \cr
                   -1/\sqrt 2 & 1/\sqrt 6 & 1/\sqrt 3 \cr
                           0 & -2/\sqrt 6 & 1/\sqrt 3 \cr
                                         } \right)\ ,  \nonumber\\
L \simeq &&\left( \matrix{\cos \theta_\ell & - \sin \theta_\ell & 
                                          \lambda_\ell\sin 2\theta_\ell \cr
                  \sin \theta_\ell  & \cos \theta_\ell & 
                                        \lambda_\ell\cos 2\theta_\ell   \cr
              -\lambda_\ell\sin 3\theta_\ell    & -\lambda_\ell\cos 3\theta_\ell & 1 \cr
                                         } \right)\ , 
\end{eqnarray}										 
with
\begin{equation}
\tan 2\theta_\ell\simeq-\sqrt{3}{\rho_\ell-\delta_\ell \over 
  2\e_\ell-\rho_\ell-\delta_\ell}\ , \qquad
   \lambda_\ell={1 \over \sqrt 2}{1 \over 3c_\ell}\xi_\ell \ .
\end{equation}
If a special condition $\delta_\ell=-\rho_\ell$ is taken,  one can obtain
 familiar relation $|L_{21}|\simeq \sin \theta_\ell \simeq\sqrt{|m_{\ell 1}/m_{\ell 2}|}$,
  which was used in refs.\cite{FX} and \cite{FTY}.
 However, this condition is  not  guaranteed in the framework of the model.
 In our following analyses,  a relation between $\delta_\ell$ and $\rho_\ell$  is  
  given  only by  the mass $m_{\ell 1}=m_e$ as seen in eq.(\ref{mass}),
 and so the value of $L_{21}$ is given arbitrary.
 The $L_{13}$ and $L_{31}$ mixings are suppressed compared with $L_{12}$ and $L_{21}$ 
 because $\l_\ell$ is  ${\cal O}(m_\m/m_\tau)$.
 On the other hand, the $L_{23}$ and $L_{32}$ mixings are almost fixed  except for
  phases as
 \begin{equation}
       |L_{23}|\simeq |L_{32}| \simeq {1\o \sqrt{2}}{m_\m\o m_\tau}\simeq 0.04  \ .
 \end{equation}

Let us turn to the neutrino sector. Assuming that the neutrinos are of
the Majorana type, the neutrino mass matrix  is 
%%%%%%%%%%%%%%%%%%%%%%%%%%%%%%%%%%%%%%%%%%
%%%%%%%%%%%% Neutrinos %%%%%%%%%%%%%%%%%%%%
%%%%%%%%%%%%%%%%%%%%%%%%%%%%%%%%%%%%%%%%%%
\begin{equation}
M_\n= {c_\n}
             \left(\matrix{1 & 0 & 0 \cr
			                0 & 1 & 0 \cr
                            0 & 0 & 1 \cr  } \right)
 +\left( \matrix{\delta_\n & 0 & 0 \cr
                 0  & \rho_\n & 0 \cr
                  0 & 0 & \e_\n\cr} \right),
\end{equation}
\noindent
 where the first matrix is $S_{3L}$ invariant one and 
  the second  is the symmetry breaking one.
  It is noted that $c_\n=0$ and  $c_\n={\cal O}(1\eV)$ were taken  in  ref.\cite{FX} and  
  in ref.\cite{FTY}, respectively.
 Since the mass matrix of the neutrino is still diagonal one in both models,
 the same  numerical results are obtained for flavor mixings.
  The neutrino masses are determined being independent of flavor mixings in this model.
  So the parameters $\delta_\n$, $\rho_\n$ and $\e_\n$ are easily constrained
  by putting $\Delta m^2_{21}=\Delta m_\odot^2$ and 
  $\Delta m_{31}^2 =\Delta m_{\rm atm}^2$.
  
%%%%%%%%%%%%%%%%%%%%%%%%%%%%%%%%%%%%%%%%%%
%%%%%%%%%%%% Mixings %%%%%%%%%%%%%%%%%%%%
%%%%%%%%%%%%%%%%%%%%%%%%%%%%%%%%%%%%%%%%%%
  
 The neutrino mixing matrix $U_{\a i}$ is determined by only $V_\ell$ of the charged leptons
  as follows:
  \begin{equation}
  U=V^\dagger_\ell=L^\dagger F^\dagger \ .
  \end{equation}
  \noindent
  Then, the relevant mixing parameters of solar neutrinos are:
  \begin{eqnarray}
  U_{e1}&=& {1\o \sqrt{2}}L_{11}^*+ {1\o \sqrt{6}} L_{21}^*     \ , \nonumber \\
  U_{e2}&=& -{1\o \sqrt{2}}L_{11}^*+ {1\o \sqrt{6}} L_{21}^*     \ , \nonumber \\
  U_{e3}&=& -{2\o \sqrt{6}} L_{21}^*     \ , 
  \label{Uei}
  \end{eqnarray}
\noindent where $|L_{11}^*|^2\simeq 1-|L_{21}|^2$ and 
   $L_{31}$ is neglected due to the suppression factor $\l_\ell$.
These mixings are determined only by $L_{21}$, which is still an unknown
parameter because there remains at least one  undetermined parameter in the mass matrix
 after fixing the charged lepton masses $m_e$, $m_\m$ and $m_\tau$ as seen in eq.(\ref{mass}).
 If the complex phases are allowed in the symmetry breaking terms, 
 there are more unknown parameters.
 On the other hand,
 the relevant mixing parameters of  atmospheric neutrinos are:
  \begin{equation}
  U_{\m 3}= -{2\o \sqrt{6}} L_{22}^*+ {1\o \sqrt{3}} L_{32}^*   \ ,  \qquad\quad
  U_{\tau 3}= -{2\o \sqrt{6}} L_{23}^*+ {1\o \sqrt{3}} L_{33}^*      \ , 
  \label{Umi}
  \end{equation}
  \noindent where $|L_{22}^*|^2= 1-|L_{21}|^2-|L_{23}|^2$ and 
  $|L_{33}^*|^2\simeq 1-|L_{32}|^2$. Thus, these mixings are determined by $L_{21}$ and
  $L_{32}$.
  
  %%%%%%%%%%%%%%%%%%%%%%%%%%%%%%%%
  %%%%%%%%  Oscillations %%%%%%%%%
  %%%%%%%%%%%%%%%%%%%%%%%%%%%%%%%%
 
  As seen in eq.(\ref{Uei}), there is a  correlation between
  $4 |U_{e1}U_{e2}^*|^2$ and  $|U_{e3}|$, both are relevant quantities for the
  oscillation probability of solar neutrinos as follows:
  \begin{equation}        
  P(\n_e\Ar \n_e) \simeq 1 - 
        4 |U_{e1} U_{e2}^*|^2 \sin^2 {\Delta m^2_{21} L \o 4 E} 
		-2 |U_{e 3}|^2 (1-|U_{e 3}|^2) \ .
  \end{equation}
  \noindent
  Since those are given in terms of $L_{21}=|L_{21}|\exp(i p)$, we can present
 allowed region on the ($|U_{e3}|$, $4 |U_{e1} U_{e2}^*|^2$) plane by
  changing  $|L_{21}|$ and the phase $p$. 
 In Fig.1, the allowed region is shown between two curves  for the case of
  $p=0^\circ$ and $p=90^\circ$.
  If $\Delta m^2_{31}\geq 9\times 10^{-4}\eV^2$, the mixing $|U_{e3}|$ is constrained 
  by  the CHOOZ experiment \cite{CHOOZ}, in which the oscillation  probability is expressed as
   \begin{equation}        
  P(\bar\n_e\Ar \bar\n_e) \simeq 1 - 
        4 |U_{e 3}|^2 (1-|U_{e 3}|^2) \sin^2 {\Delta m^2_{31} L \o 4 E} \ .
  \end{equation}
  \noindent
 For example, the CHOOZ result gave  constraints  $\sin^2 2\th_{\rm CHOOZ}\leq \{0.12,0.20,0.75\}$
  for  $\Delta m^2_{31}\simeq \{5,2,1\}\times 10^{-3} \eV^2$,
   which correspond to $|U_{e 3}|\leq \{0.18,0.23,0.50\}$, respectively.
   Thus, the constraint of $|U_{e 3}|$ is strongly depend on  $\Delta m^2_{31}$.
   This fact is advantage of testing the model because the solar neutrino data
   determines the best fit  of  ($|U_{e3}|$, $4 |U_{e1} U_{e2}^*|^2$)
    being independent of  $\Delta m^2_{31}$.
	We show the best-fit points in the three flavor analyses in ref.\cite{Osland}
	by four black points for each $|U_{e3}|$.
	 If these best-fit points are reliable,
	we can find in Fig.1 that  $|U_{e3}|\simeq 0.3\sim 0.4$ is favored in the present model, 
	which means $\Delta m^2_{31}\leq 10^{-3}\eV^2$ in the CHOOZ data.
	LBL experiments can test this  prediction in the future.  
	  
	   However, it is important to comment on the allowed region at $95\%$ C.L.
	    in ref.\cite{Osland}, which  completely covers our
	  predicted one. Thus, the prediction of  $|U_{e3}|$ in the model is less predictive
	   at the present. Moreover, for  $|U_{e3}|=0.0, 0.2$ and $0.4$, 
	    the values of $\chi^2_{\rm min}$ are $3.6$, $4.2$  and $5.6$, respectively.
		  This means that smaller values of $|U_{e3}|$ is prefered.
	Therefore, we need three flavor analyses of the solar neutrino oscillation 
	including new data of
	energy spectrum in Super-Kamiokande \cite{SKamsolar} \cite{Bahcall}. 
	
 %%%%%%%%%%%%%%%%%%%%%%%%%%%%%%%%%%
 %%%%%%%%%%%%  LBL  %%%%%%%%%%%%%%%
 %%%%%%%%%%%%%%%%%%%%%%%%%%%%%%%%%%	
	
	 LBL experiments provide an important test of the model
	because $|U_{e3}|$ is a key ingredient for the $\n_\m\Ar \n_e$ oscillation
	as follows:
	\begin{equation}        
  P(\n_\m\Ar \n_e) \simeq 
        4 |U_{e3} U_{\m 3}^*|^2 \sin^2 {\Delta m^2_{31} L \o 4 E} 
		\simeq 
        {8\o 3}|U_{e3}|^2 \sin^2 {\Delta m^2_{31} L \o 4 E} \ ,
  \label{LBL}
  \end{equation}
  \noindent
  where $|U_{\m 3}|=2/\sqrt{6}$ is put and  the small $CP$ violating term is neglected.
  Then, the expected maximal oscillation probability is  $0.06$
  for $\Delta m^2_{31}\simeq 5\times 10^{-3} \eV^2$ at K2K \cite{Tanimoto} 
   due to the constraint of the CHOOZ result\cite{CHOOZ}.
    
 %%%%%%%%%%%%%%%%%%%%%%%%%%%%%%%%%%%%%%%%%%%%%%%%%%%%%%%%%%%%%%%%%%%%%%%%%%%%%%%%  
 %%%%%%%%%%%%%  Atmospheric Neutrinos  %%%%%%%%%%%%%%%%%%%%%%%%%%%%%%%%%%%%%%%%%%
 %%%%%%%%%%%%%%%%%%%%%%%%%%%%%%%%%%%%%%%%%%%%%%%%%%%%%%%%%%%%%%%%%%%%%%%%%%%%%%%%
  Let us consider  the case of the atmospheric neutrino.
   By using  $L_{32}\simeq 0.04 \exp(i q)$, we can estimate    
    $4 |U_{\m 3}  U_{\tau 3}^*|^2$  as a function of $|U_{e3}|$, 
	where $q=0^\circ \sim 180^\circ$ and
	$L_{23}$ is determined by unitarity.
 We present allowed region on the ($|U_{e3}|$, $4 |U_{\m3} U_{\tau 3}^*|^2$) plane 
 in Fig.2, where the allowed region is shown between two curves  for the case of
 the phase $q=0^\circ$ and $q=180^\circ$.
 It is found that $4 |U_{\m 3}  U_{\tau 3}^*|^2$ could deviates significantly  
 from $8/9$, which was presented in refs.\cite{FX}\cite{FTY}, 
 due to the symmetry breaking terms in the model.
 The LBL experiments of $\n_\m\Ar \n_\tau$ and $\n_\m\Ar \n_e$
   enable to test our predictions in the future. 
  
 %%%%%%%%%%%%%%%%%%%%%%%%%%%%%%%%%%%%%%%%%%%%%%%%%%%%%%%%%%%%%%%%%%%%%%%%%%%%%%%%%
  The summary is given as follows.
  The mass matrix model based on the democratic form
  has been  investigated focusing on the recent data of solar neutrinos and atmospheric neutrinos.
  If the solar neutrino deficit is due to  the vacuum oscillation as suggested 
  by recent Super-Kamiokande data,  nearly bi-maximal mixings are needed 
   in order to explain  both solar and 
  atmospheric neutrino deficit. In the present model, 
   those large mixings are derived from the charged lepton sector
  while the neutrino mass matrix is diagonal one.
  It is remarked that the symmetry breaking term of the charged lepton mass matrix
  is very important to predict mixing angles.
  The model will be tested by the precise determination of the mixings and
  $\Delta m^2$ in the solar neutrinos, atmospheric neutrinos and
    LBL experiments in the near future.
	\vskip 1 cm
  
  %%%%%%%%%%%%%%%%%%%%%%%%%%%%%%%%%%%%%%%%%%%%%%%%%%%%%%%%%%%%%%%%%%%%%%%%%%%%%%%%%
  %%%%%%%%%%%%%%%%%%%%%%%%%%%%%%%%%%%%%%%%%%%%%%%%%%%%%%%%%%%%%%%%%%%%%%%%%%%%%%%%% 
  
 I thank A. Smirnov for the quantitative  discussion of 
                                vacuum oscillations of  solar neutrinos.
  I also thank Y. Koide for the discussion of the model.
 This research is  supported by the Grant-in-Aid for Science Research,
 Ministry of Education, Science and Culture, Japan(No.10140218, No.10640274).  
   
\newpage
%%%%%%%%%%%%%%%%%%%%%%%%%%%%%%%%%%%%%%%%%%%%%%%%%%%%%%%%%%%%%%%%%%%%%%%%%%%%%%%%%%%%%%%%

 \newpage

%%%%%%%%%%%%%%%  Figure 1  %%%%%%%%%%%%%%%%%
\begin{figure}
\epsfxsize=12 cm
\centerline{\epsfbox{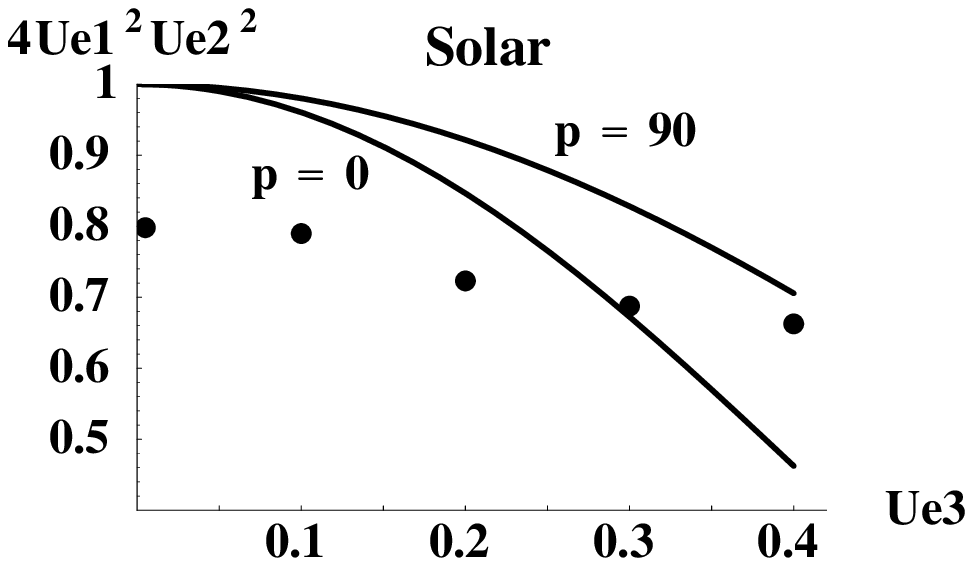}}
%\caption{}
	 \end{figure}
Fig. 1: Allowed region on ($|U_{e3}|$, $4 |U_{e1} U_{e2}^*|^2$) plane,
  which is the one between two curves for $p=0^\circ$ and $p=90^\circ$.
  Black points denote the best fits of  solar neutrinos 
  in the three flavor analyses in ref.\cite{Osland} for each $|U_{e3}|$.
  For  $|U_{e3}|=0.0, 0.2$ and $0.4$, 
	    the values of $\chi^2_{\rm min}$ are $3.6$, $4.2$  and $5.6$, respectively.
  The allowed region at $95\%$ C.L. completely covers our predicted one. 
   
%%%%%%%%%%%%%%%%%%%%%%%%%%%%%%%%%%%%%%%%%%%%
%%%%%%%%%%%%%%%  Figure 2  %%%%%%%%%%%%%%%%%
%%%%%%%%%%%%%%%%%%%%%%%%%%%%%%%%%%%%%%%%%%%%
\newpage
\begin{figure}
\epsfxsize=12 cm
\centerline{\epsfbox{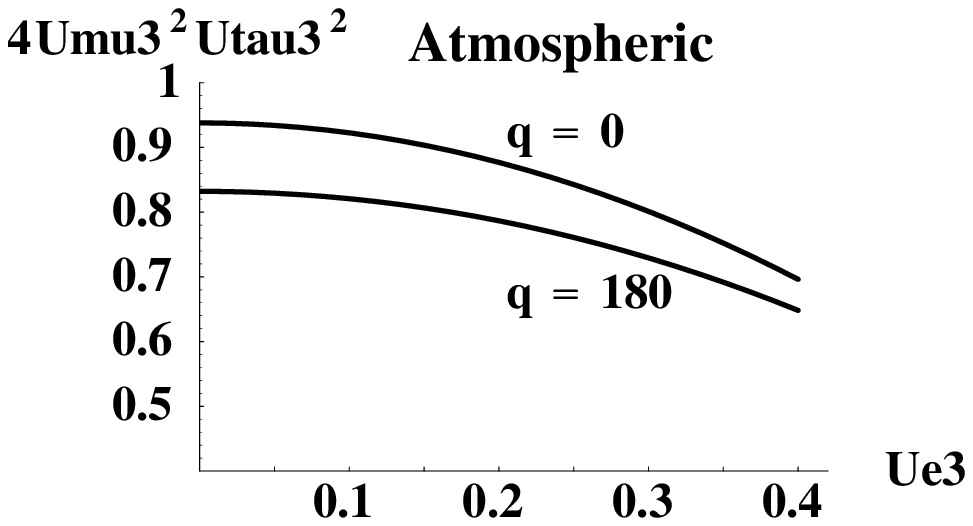}}
%\caption{}
\end{figure}
Fig. 2: Allowed region on the ($|U_{e3}|$, $4 |U_{\m3} U_{\tau3}^*|^2$) plane
       which is the one between two curves  for  $q=0^\circ$ and $q=180^\circ$.
        
%%%%%%%%%%%%%%%%%%%%%%%%%%%%%%%%%%%%%%%%%%%%% 
\end{document}